\documentclass[journal]{IEEEtran}
\usepackage[version=3]{mhchem} 
\usepackage{caption}

\usepackage{wasysym}
\usepackage{gensymb}
\usepackage{siunitx}
\usepackage{amsmath,amssymb}
\usepackage[version=3]{mhchem}
\usepackage{textgreek}
\usepackage{bm}
\usepackage[T1]{fontenc}
\usepackage [english]{babel}
\usepackage [autostyle, english = american]{csquotes}
\MakeOuterQuote{"}
\usepackage{subfiles}
\usepackage{fancyvrb}
\usepackage{blindtext}
\usepackage{caption}
\usepackage{graphicx}
\usepackage{xr}
\usepackage{color,soul}

\AtBeginEnvironment{thebibliography}{\linespread{1}\selectfont}
\usepackage{hyperref}
\usepackage{cleveref}
\usepackage{booktabs}
\usepackage[table,xcdraw]{xcolor}

\usepackage[moderate]{savetrees}
\begin{document}

\title{The carbon cost of a PhD in solar cell materials}


\author{\large Rachel Woods-Robinson\textsuperscript{*†}
\thanks{* University of California and Lawrence Berkeley National Laboratory, Berkeley, CA, 94720 USA}
\thanks{† National Renewable Energy Laboratory, Golden, CO 80401, USA}
}


\maketitle

\begin{abstract}

Scientific research can bring about important progress in developing renewable energy technology such as solar cells. However, in addition to positive impact, research also comes at a cost, and one such cost is the \ce{CO2} footprint associated with doing research and producing technologies. Here, as an example of such a cost, I estimate the associated greenhouse gas (GHG) emissions associated with my PhD, in which I researched computational and experimental design of thin film materials for photovoltaics. I consider supercomputer power, synchrotron beam time, laboratory synthesis and characterization, travel associated with research, and personal emissions over the course of my PhD. This analysis demonstrates that doing science increases my carbon footprint to much higher than an average American. Because our consumption is so high, scientists have a moral imperative to reduce our own emissions in our laboratories and in the technologies we create.

\end{abstract}

\IEEEpeerreviewmaketitle

\section{Introduction}

The intention of much of our research within the science and engineering community is to develop and optimize materials necessary for our transition to a renewable energy society, and ultimately to contribute to a reduction of \ce{CO2} emissions. However, this research comes at a cost: time, resources, energy, and ultimately \ce{CO2} and other greenhouse gases (GHGs). As a scientist, I believe it is important to assess how our actions and our research activities impact society, and to acknowledge that research can have both positive and negative impacts. With an awareness of our impact on society, we can then take action and make changes to ensure that our science is done a way that minimizes GHG footprint as well as minimizes other forms of harm. In order to reach our renewable energy goals, scientists need to think critically about how we are using energy and resources within the work we do.

In this analysis, I focus on a personal case study to quantify GHG emissions due to my own research activities and contrast them with emissions of other members of society. Specifically, I seek to estimate the GHG emissions incurred by my PhD research studying computational and experimental materials discovery of inorganic semiconductors for solar energy applications. I find that the primary energy-intensive aspects of my research have been (1) supercomputer time, (2) synchrotron beam time, (3) laboratory synthesis and characterization, and (4) travel associated with research and conferences, and I include an estimate of my personal GHG emissions for comparison. I also estimate a few alternate scenarios as a way of contextualizing these emissions, and compare these scenarios to my actual emissions by source.

I acknowledge that I have limited access to granular data associated with my research and do not expect these results to be perfectly accurate; rather, my intention is to quantify the order of magnitude of the GHG emissions and to open a dialogue within the scientific community about how much energy and \ce{CO2} is used doing renewable energy research. Before getting into the analysis, however, it must be acknowledged that Big Oil coined the term "carbon footprint" and for decades has championed personal carbon accounting to shift blame to individuals and evade changing hostile business practices.\cite{solnit2021big} I still see such analysis as a useful tool for self-assessment and as a means to scale up emissions across an individual industry, but this cannot blind-sight us from the urgent need for massive systemic change.

\section{Methodology}

In this analysis, various components of \ce{CO2} emissions related to my five-year PhD are estimated as described in the following sections. Throughout this discussion, there are two key metrics. The first is energy use. Most of this is electricity, but there is also energy associated with heating and transportation. When possible I will focus on energy \textit{consumption}, rather than production or capacity. I will report energy in units of kWh or MWh. Some of this analysis uses explicitly quantified energy consumption metrics, and in other areas this has been estimated. The second is associated GHG emissions. Most of the emissions discussed here are from \ce{CO2}, but other GHG include methane (\ce{CH4}) and nitrous oxide (\ce{N2}), among others. The impact of other GHGs is quantified in terms of units of \ce{CO2}-eq., which is derived from their global warming potential (GWP). The metrics of energy consumption and \ce{CO2}-eq. are connected by the \ce{CO2}-eq. emissions factor for a given energy source or mix, which is measured in units of mT \ce{CO2}-eq. per MWh or g \ce{CO2}-eq. per kWh. Accurate accounting of carbon emissions associated with a given unit energy is nontrivial; attribution depends on whether one is considering the physical origin of the energy ("locational method") or "market-based" power purchasing agreements, among other factors.\cite{stechemesser2012carbon, brander2018creative} For the purposes of this simple assessment, we will assume the market-based emission factors as reported in \autoref{tab:emission-factors-source} and \autoref{tab:emission-factors-location} from various references,\cite{epa2019greenhouse,kotowicz2018analysis,pehl2017understanding,ultra2021ultra,xcel2019carbon,caiso2021today,sgip2021greenhouse,epa2019greenhouse,letemps2009switzerland} though we acknowledge that considering consumption-based economic metrics alone can lead to misallocation of emissions.

\section{Results and Discussion}


\begin{table}[]
\centering
\caption{GHG emission factors of various electricity sources used in this analysis.}
\label{tab:emission-factors-source}
\resizebox{0.45\textwidth}{!}{%
\begin{tabular}{@{}ccc@{}}
\toprule
\textbf{Source} & \textbf{Emissions factor (g \ce{CO2}-eq / kWh)} &  \textbf{Ref.} \\ \midrule

Coal                           & 990--1,100 &  \cite{epa2019greenhouse}       \\
Combined cycle gas power plant & 330--449   &  \cite{kotowicz2018analysis}       \\
Wind                           & 4--14        &  \cite{pehl2017understanding}        \\
PV                             & 5--50        &  \cite{pehl2017understanding}        \\
Ultra low-carbon PV            & 5           &  \cite{ultra2021ultra}       \\
Nuclear power plant            & 3.5         &  \cite{pehl2017understanding}        \\ 

\bottomrule
\end{tabular}
}
\end{table}

\begin{table}[]
\centering
\caption{GHG emission factors for the grid at a particular location.}
\label{tab:emission-factors-location}
\resizebox{0.45\textwidth}{!}{%
\begin{tabular}{@{}cccc@{}}
\toprule

\textbf{Location} & \textbf{Emissions factor (g \ce{CO2}-eq / kWh)} & \textbf{Assumptions}   & \textbf{Ref.} \\ \midrule

NREL (Xcel CO)                 & 544         & 2019 levels   & Xcel\cite{xcel2019carbon}    \\
California grid                & 220         & 2018 levels   & CAISO\cite{caiso2021today,sgip2021greenhouse}    \\
SLAC before Nov. 2019      & 1100                                       & 2011 Navajo coal plant & EPA\cite{epa2019greenhouse}          \\
Switzerland                    & 100         & 2009 estimate & Chevalley\cite{letemps2009switzerland} \\

\bottomrule
\end{tabular}
}
\end{table}

\begin{figure}
    \centering
    \includegraphics[width=0.5\textwidth]{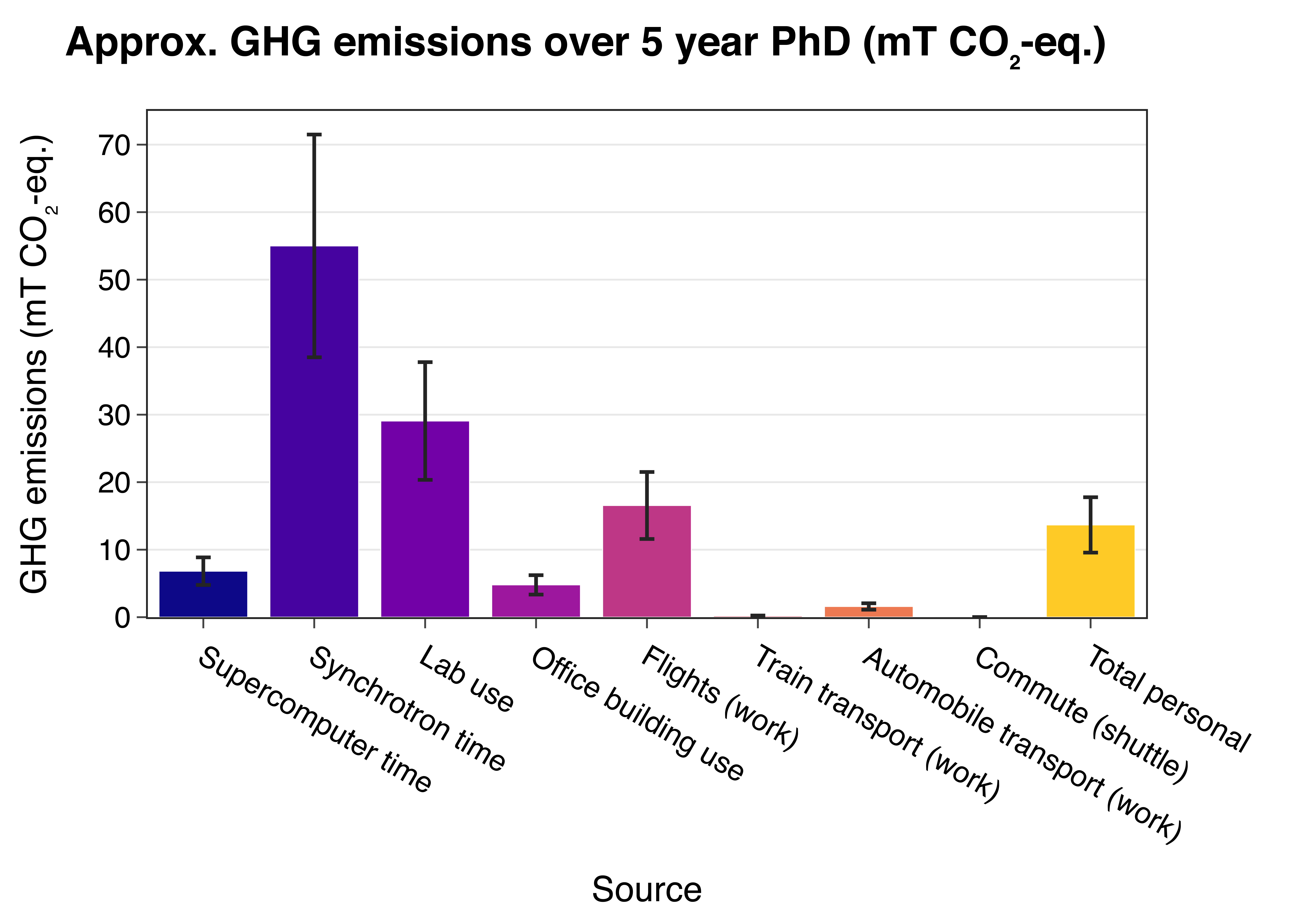}
    \caption[Estimate of GHG emissions over the course of my five-year PhD]{Estimate of GHG emissions over the course of my five-year PhD, as approximated in the text.}
    \label{fig:PhD-emissions}
\end{figure}

\subsection{Research-related emissions}

\subsubsection{Supercomputer emissions}

The increase of computational capabilities have enabled a wide array of new research frontiers. In addition to materials discovery, supercomputers can be used for complex climate change simulations, and to solve other difficult problems related to our sustainable development. For example, in my research I use supercomputers and \textit{ab inition} methods to simulate the structure and properties of new contact materials for solar energy applications, and increased computer power enables high-throughput capabilities so I can screen through large quantities of materials.\cite{jain2013commentary} However, this creates a new dilemma. The more complex a simulation is, the more energy it uses, and currently supercomputers and data serves are incredible energy intensive. If our computers and data centers are not powered by renewable energy, using them for research may come at a steep \ce{CO2} cost. Although many of us are working to combat the climate crisis, rarely do we discuss our own carbon footprint from computing.

The computational research in my PhD was performed using the supercomputer Cori, which is one of the National Energy Research Scientific Computing Center (NERSC) machines. My research at NERSC has used approximately 8 million CPU hours over five years.\cite{nersc2021iris} This is very likely at the low end for my computational materials scientists, since I only spent about half of my research time doing calculations. For 1,126 of the 1,527 jobs I ran in the year 2020, 1.05 million CPU hours used a total of 2,431 kWh of electricity. These jobs are mostly density functional theory calculations using \texttt{VASP}, though I also run \texttt{MCSQS} and some bash scripts directly. This corresponds to approximately 0.0024 kWh associated with each CPU hour for my research. Assuming a similar energy use rate applies to all computational jobs I have run during my PhD, this approximation allows me to estimate the energy use associated with other years of computation time, and approximate that 8,169,603 CPU hours corresponds to $\sim$18.83 MWh over five years and $\sim$3.766 MWh per year (less energy per CPU hour than the average NERSC user). Assuming the emissions factor of the California grid from the California Independent System Operator (CAISO),\cite{caiso2021today,sgip2021greenhouse} reported in \autoref{tab:emission-factors-location} as $\sim$0.362 mT \ce{CO2}-eq./MWh,\footnote{This is an estimate from CAISO based on granular data from the electricity grid.\cite{caiso2021today}} this amounts to \textbf{6--8 mT \ce{CO2}-eq.} across my five year PhD due to supercomputer time.

\subsubsection{Synchrotron beam time emissions}

Experiments at synchrotron particle accelerators can enable scientists to probe fundamental properties of materials, and in my research I used the Stanford Synchrotron Radiation Lightsource (SSRL) to perform X-ray diffraction and X-ray absorption on thin film materials for solar cells. To my knowledge, SSRL does not share data about facility and beam line energy and GHG usage, so here I will estimate emissions during my synchrotron experiments based on macroscopic energy data and power-purchasing agreements. The synchrotron at SSRL uses a $\sim$2.5 MW X-ray radiation source called SPEAR,\footnote{Energy information about SPEAR is from personal correspondence with Dr. David Chassin and Cezary Jach.} and operates around the clock for 10.5 months/year (7,560 hr/year). This corresponds to a typical energy use of 2.5 MW $\times$ 7,560 hr/year = 18.9 MWh/year. The SSRL synchrotron has 30 stations and 14 beamlines.\cite{ssrl2021experimental} I will assume as a user that I am responsible for energy associated with an entire beam line, although it is sometimes shared during operation so this estimate is an upper limit. During my PhD, I was granted and was responsible for 790 hours of beam time, although I was only present at about 50\% of these beam times. Thus, assuming an error bar of $\pm$25\% to account for uncertainties in this estimate, I used approximately 50 MWh total of synchrotron beam energy during my PhD.

Until November 18, 2019, SSRL's power-purchasing agreement with the Western Area Power Administration (WAPA) marketed power from the Navajo Generating Station in Arizona. This was a coal plant (that has since been decommissioned),\cite{epa2019greenhouse} and since all of my beam time was performed before this date, "economically" all of the energy I used came from coal. From Stanford's Western Interconnection Data Analytics Project (WIDAP) database,\cite{stanford2018widap, grueneich2018western} emissions from this power plant were approximately 1.1 mT \ce{CO2}/MWh in 2016--2018 (as well as 2 mT \ce{NO_$x$} and 1.6 mT \ce{SO_$x$}). Thus, my 50 MWh energy consumption corresponds to approximately \textbf{55 $\pm$ 14 mT \ce{CO2}-eq.} from synchrotron beam time.\footnote{A more accurate estimate of synchrotron emissions would use real-time location-based data, however such data was not available at the time of this assessment.}

Although synchrontron emissions reported here are associated with coal, the power purchasing agreement has changed and currently power for the synchrotron is 100\% from wind power. According to the 2018 Intergovernmental Panel on Climate Change (IPCC) report, the \ce{CO2}-eq. emissions associated with wind power range from 8--20 g / kWh (0.014 $\pm$ 0.006 mT \ce{CO2}-eq. / MWh).\cite{ipcc2018global} Therefore, assuming this range, if I was to start my PhD today and perform the same amount of work at SSRL my associated emissions would be about two orders of magnitude lower: approximately 0.7 $\pm$ 0.3 mT \ce{CO2}-eq.

\begin{figure}
    \centering
    \includegraphics[width=0.55\textwidth]{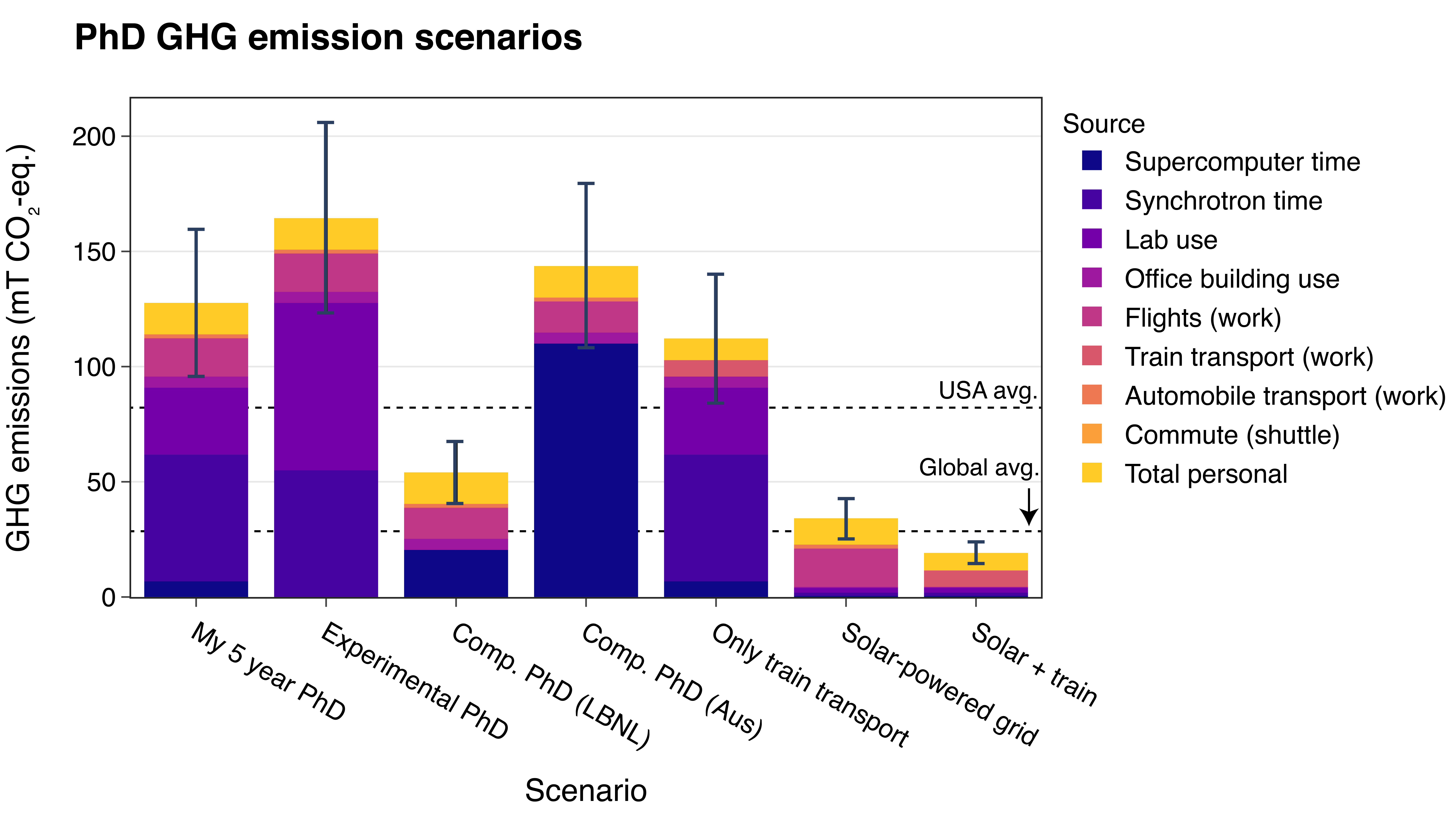}
    \caption[Alternative scenarios of emissions during my PhD]{Alternative scenarios of emissions during my PhD.}
    \label{fig:emissions-scenarios}
\end{figure}

\subsubsection{Emissions associated with laboratory work}

Estimating emissions associated with laboratory work is nontrivial, since I have used a variety of tools at a variety of institutions, and associated energy contributions spans not just the time I used the tool but the energy required for keeping the facility running, the energy source of the particular facility, the specific experiment run, and the number of users, among other factors. LBNL collects and shares macroscopic data about its facility energy consumption, from 2014 onward,\cite{lbnl2021energy} however I was unable to find similar aggregated data for the other labs I worked at. Therefore I will make a rough assumption here that LBNL energy use data is representative of my total lab use. Total energy use at LBNL in 2019 was 142,230,864 kWh, and dry labs used $\sim$16\% of this, i.e. a total of 18,490 MWh/year across the entire lab. Using a variety of assumptions based on number of experimentalists at LBNL, I estimate I am responsible for $\sim$15 MWh/year from dry lab usage, or $\sim$75 MWh over five years. To roughly account for the variability in the energy source at the different labs I worked at based on their electricity grids (see \autoref{tab:emission-factors-location}), together, this amounts to a weighted average of 0.3875 mT \ce{CO2}-eq. / MWh, and $\sim$5 mT \ce{CO2}-eq. / year from dry labs. Since this quantity is based on a series of assumptions, I will assume an error propagation of $\pm$50\%. In total this amounts to \textbf{$\sim$25$\pm$12 mT \ce{CO2}-eq.} over my PhD from dry lab use.

\subsubsection{Office building emissions}

Using a similar framework as above, I estimate the associated emissions with working in an office using LBNL's aggregated energy use data.\cite{lbnl2021energy} In 2019, office work contributed towards 9\% of energy use at LBNL, i.e. 12,800 MWh/year. Assuming emission factors from \autoref{tab:emission-factors-location} and \autoref{tab:emission-factors-source}, as well as similar attribution assumptions as discussed for laboratory emissions, this corresponds to approximately 1.74 mT at NREL, 2.99 mT at LBNL, and 0.053 mT at EPFL. In total, I can attribute \textbf{$\sim$4.8 mT \ce{CO2}-eq.} to my use of office buildings during my PhD.

\subsection{Work-related travel and transport emissions}

Work-related travel during my PhD included travel for collaborative research projects, summer schools and workshops, and international conferences. In total, I traveled 134,159 miles (215,908 km) by plane, and 102,816 of these miles (165,464 km) was for work-related travel. Emission factors for commercial aviation vary depending on the specific aircraft, its weight, its engine, weather, and number of passengers; reported literature emission values range from approximately 90--100 g \ce{CO2}-eq. per passenger per km.\cite{carbon2021aviation, graver2020}. Assuming a more conservative estimate of 100 g \ce{CO2}-eq. per passenger per km, I estimate I am accountable for approximately 21.6 mT \ce{CO2}-eq. emissions total, and thus \textbf{16.5 mT \ce{CO2}-eq.} from my work-related travel. In addition, I traveled 2859 miles (4601 km) by train. Assuming train travel emissions of approximately 41 g per km per passenger,\cite{ritchie2020co2} my associated train travel emissions are approximately \textbf{0.19 mT \ce{CO2}-eq.}. I do not own a car but have traveled approximately 2400 miles (3860 km) by car for work. Assuming automobile emissions of 411 g of\ce{CO2}-eq. per mile from the EPA's GHG calculator from 2019,\cite{epa2019greenhouse} my associated automobile emissions from travel are quite low at \textbf{0.99 mT \ce{CO2}-eq}. I commuted to work by bicycle, with the occasional $\sim$1 mile shuttle ride to LBNL or bus ride to NREL. Approximating one shuttle ride every week for five years amounts to 260 miles of bus transport, and, using the automobile emissions factor as an upper limit, this results in \textbf{0.11 mT \ce{CO2}-eq} from bus transport.

\subsection{Personal emissions}

Here, I use the EPA carbon footprint calculator to estimate my personal carbon footprint.\cite{epa2021carbon} As inputs, I used the data from my electricity and gas bills in Berkeley, CA, estimating $\sim$100 kWh/month for electricity and $\sim$14 therms/month for gas. According to this estimate, I am responsible personally for 3,102 lbs \ce{CO2}-eq. per year, i.e. 1.4 mT \ce{CO2}-eq. / year and 7 mT \ce{CO2}-eq. over the course of my five year PhD. Accounting for the associated emissions from personal flight, train, bus, and automobile transportation, this adds approximately 5.04, 0.13, 0.04, and 0.88 mT, respectively, with a total of 6.1 mT. Thus, my estimate of personal emissions amounts to \textbf{13.1 mT \ce{CO2}-eq.} over the course my PhD. For comparison, the emissions of an average American are $\sim$16.2 mT \ce{CO2}-eq. per year. \cite{ritchie2020co2} I expect in reality my personal emissions are less than this because of my vegetarian diet, because I do not own a car, and because I live in a location that does not require much heating or cooling.

\section{Discussion}

In \autoref{fig:PhD-emissions}, I plot all of the emissions that have been estimated previously. The biggest sources of emissions during my PhD has been from laboratory research and computation rather than from flights. Synchrotron beam time accounts for the largest portion of my personal emissions. This is due primarily to the power purchasing agreement at SSRL that was primarily coal power, though this has since been decommissioned. These cumulative emissions are also plotted in the first bar of  \autoref{fig:emissions-scenarios}, showing total emissions of $\sim$128 mT \ce{CO2}-eq. over my five year PhD. This is approximately 63\% higher than the average per capita US emissions over five years of $\sim$81 mT \ce{CO2}-eq. ($\sim$16.2 mT \ce{CO2}-eq. annually).\cite{ritchie2020co2} For comparison, the global average is 4.8, or $\sim$24 mT \ce{CO2}-eq., and this is shown in the plot as well.

Here, I estimate a few alternate scenarios as a way of contextualizing these emissions, and compare these scenarios to my actual emissions by source in \autoref{fig:emissions-scenarios}. First, I estimate what these emissions would be have been if my PhD had been solely experimental or solely computational research. For an experimental PhD, I multiply the lab-based emissions by a factor of 2.5, and remove computational emissions. For a purely computational PhD, I use an estimate based on the total allocated time of my group. This reduces emissions by nearly half; however, it is important to note that more complex computational simulations, e.g. machine learning and artificial intelligence models, can use significantly more energy and thus would be attributed to far more emissions from supercomputer time. Additionally, this number is very dependent on the emissions factor of the supercomputer. For example, it has been shown that the astronomy community in Australia on average contributes to 22 mT \ce{CO2}-eq. per person per year, or 110 mT \ce{CO2}-eq. over five years.\cite{stevens2020imperative} "Computational PhD (Aus. approx.)" is plotted assuming these emissions (with all other sources unchanged), showing a significantly higher contribution from compute time.



Next, I estimate alternate scenarios given different emissions factors. I estimate emissions if, rather than flying and driving, all of my transportation had been on trains. This somewhat reduces transportation-associated emissions, although other research associated emissions still keep the total significantly higher than the US average. I then estimate the impact of switching to a 100\% solar-powered electrical grid. This would impact nearly every emissions source, bringing my impact down to nearly half of the US average, though still higher than the global average with most of my emissions due to flying. And lastly, I estimate the impact of a solar grid and only traveling by train. This would cut my emissions by nearly 5 times, bringing them to approximately the global average. The takeaway here is that, regardless of research task, switching to a renewable grid and offsetting travel to use trains rather than planes (or traveling less frequently) is extremely impactful on the carbon footprint of a researcher.


\section{Conclusion}

In this paper I have estimated my carbon footprint during my five-year PhD, showing that my total emissions are $\sim$128 mT \ce{CO2}-eq. or $\sim$25.6 mT \ce{CO2}-eq. annually. This is approximately 63\% higher than the average per capita US emissions over five years of $\sim$16.2 mT \ce{CO2}-eq. annually\cite{ritchie2020co2} and over 25 times higher than average global per capita emissions. Switching to a less carbon-intensive electricity grid and cleaner methods of transportation should lower research-related emissions. However, to get to our goals of net-zero carbon emission by 2050 as set by the IPCC, global per capita emissions must decrease to less than 2.5 mT \ce{CO2}-eq. annually. The decreases from an entirely solar-powered grid and only train transportation are still not enough to reach this 2050 net-zero target! The key takeaway is that research and activities associated with a materials science career significantly increase my carbon footprint beyond that of a typical citizen. Our research has a significant carbon cost.

I expect that this finding applies generally to scientists across the US, and around the world, and I expect it to be an underestimate. This is because this assessment does not include the carbon impact that the technologies we produce will have, nor the damage from mineral extraction and waste products such technologies may induce, though such life cycle assessments are also critical. It is optimistic that switching to renewable power sources within our laboratories and decreasing travel drastically reduces emissions, and our energy grids are becoming more renewable overtime, but this is not happening fast enough. Using energy and \ce{CO2} to address our energy \ce{CO2} challenges is indeed necessary, and our research is essential to making progress in the energy transition. However, because our consumption is so high, scientists across the planet have a moral imperative to take action and to demand that the energy used for our research comes from renewable sources.

\section{Acknowledgments}

I would like to acknowledge Dr. Sridutt Bhalachandra and Dr. Norman Bourassa who helped query NERSC supercomputer data, as well as Dr. David Chassin and Cezary Jach who assisted with the SSRL synchrotron energy and \ce{CO2} approximation. Dr. Soren Scott provided inspiration for this analysis.\cite{scott2019isotope} Special thanks to Kyle Graycar, Dr. Sebastian Husein, Dr. Matthew Horton, Dr. Kristin A. Persson, and Dr. Andriy Zakutayev for reviewing this text in various capacities. I would also like to acknowledge and honor people across the planet working together to combat climate change and environmental destruction.

\ifCLASSOPTIONcaptionsoff
  \newpage
\fi




\bibliographystyle{IEEEtran}
\bibliography{refs.bib}

\end{document}